\documentstyle[12pt,epsf]{article}
\begin{document}
\begin{titlepage}
%\begin{flushright}
%LPTENS 95/37\par\noindent
%hep-th/9508111
%\end{flushright}

\begin{center}
{\Large\bf An $SU(2)$ Analog of the  Azbel--Hofstadter Hamiltonian}

\vskip2truecm

E. G. Floratos$^{*}$\footnote{e-mail: floratos@cyclades.nrcps.ariadne-t.gr.
On leave of absence from Physics Department, University of Crete.}
 and S. Nicolis$^{**}$\footnote{e-mail: nicolis@celfi.phys.univ-tours.fr}

\vskip1truecm 

{\sl CNRS--Laboratoire de Physique Th\'eorique 
de l'Ecole Normale Sup\'erieure\footnote{Unit\'e propre du CNRS (UPR 701) associ\'ee \`a l'ENS et \`a l'Universit\'e Paris-Sud.}\\
24 rue Lhomond, 75231 Paris Cedex 05, France}

\vskip2truecm

$^{*}${\sl I. N. P. , NRCPS ``Demokritos''\\
15310 Aghia Paraskevi, Athens, Greece} 

\vskip1truecm

$^{**}${\sl CNRS--Laboratoire de Math\'ematiques et 
 Physique Th\'eorique\footnote{Unit\'e Propre d'Enseignement et de Recherche 
 (UPRES A 6083) associ\'ee \`a l'Universit\'e de Tours.} \\
 Universit\'e de Tours,
Parc Grandmont, 37200 Tours, France}

\end{center}

\vskip1truecm

\begin{abstract}
Motivated by recent findings due to Wiegmann and Zabrodin, Faddeev
 and Kashaev concerning the appearence of the quantum $U_q(sl(2))$ symmetry
in the problem of a Bloch electron on a two-dimensional magnetic lattice, we
introduce a modification of the tight binding Azbel--Hofstadter 
Hamiltonian that is a specific spin$-S$ Euler top and can be 
considered as its ``classical'' analog. The eigenvalue problem for the 
proposed model, in the coherent
state representation, is described by the $S-$gap Lam\'e equation and, thus, 
is completely solvable. We observe a striking similarity between the shapes 
of the spectra of the two models for various values of the spin $S$.  
\end{abstract}

\end{titlepage}
\section{Introduction}

The quantum mechanics of free electrons on two dimensional lattices, in the 
presence of a homogeneous and transverse magnetic field,  (``magnetic lattices'')
leads to the discovery of a host of physically and mathematically 
fascinating problems. 

This subject has a long history, starting with the pioneering papers of Harper, 
Azbel, Zak and Chambers, Hofstadter and Wannier~\cite{harper,azbel,zak,chambers,
hofst,wannier}. With the 
discovery of the quantum Hall effects~\cite{qhe,advance,review}, a large number of 
very interesting theoretical papers appeared, which deal with the quantum mechanical 
explanation of the Hall conductivity plateaus~\cite{laughlin,scaling}.  

More recently,  in  the work 
of Wiegmann and Zabrodin~\cite{wz}, it was  found that the eigenvalue problem for the 
the  Hamiltonian of the one electron lattice problem (for rational magnetic flux per 
plaquette),
 henceforth called the Azbel--Hofstadter (AH) Hamiltonian, 
 can be written as a $q-$difference quadratic equation, 
using the quantum group $U_q(sl(2))$, known as the ``Jimbo''
deformation of $SU(2)$, which appears naturally in this context.  Their analysis
leads to the algebraic Bethe Ansatz equations for the roots of its polynomial
solutions, at a particular point  of the Brillouin zone of the square lattice. 
Subsequently, Faddeev and Kashaev proved that this symmetry exists at 
{\em all} points of the Brillouin zone for square and triangular (anisotropic, 
in general) lattices and they provided the 
corresponding Bethe Ansatz equations~\cite{fadkas}. 

However, it has not yet been possible to {\em solve} the Bethe {\em Ansatz} 
equations thus obtained. Direct numerical solutions were found, in certain 
cases by Hatsugai, Kohmoto and Wu~\cite{kohmoto}; but the necessity of a 
systematic approximation scheme remains an open issue.

In this paper we  propose a modification of the AH Hamiltonian through a specific 
spin$-S$ Euler top, which has the merit of being completely solvable. Indeed, 
in the coherent state representation, the eigenvalue problem for this Euler 
top is described by an $S-$gap Lam\'e equation. Numerical comparison of the 
spectra of the two models reveals a striking similarity for their shapes. 
  
In section~\ref{notation} we establish a uniform notation, recalling, 
at the same time, the salient results of ref.~\cite{wz,fadkas}.  
In section~\ref{slq2C}, we write the AH Hamiltonian 
 in terms of the generators of the Cartesian $q-$deformation~\cite{wfzhed} of $SU(2)$ 
and set the stage 
for a model Hamiltonian, ${\cal H}_1$, which is that of an ``Euler top'' under 
the ordinary $SU(2)$ group. 
In the process we will establish connections between the two different 
$q-$deformations of $SU(2)$, a problem that is interesting in its own right. 

In section~\ref{su2qsym} we discuss the eigenvalue problem for the AH 
Hamiltonian and present the recursion relations for the eigenvectors and 
the eigenvalue equation in a compact $2\times 2$ matrix form. The case 
$E=0$ admits an explicit solution~\cite{wz,kohmoto}.

In section~\ref{symmetries} we explore the  symmetries of the classical 
Hamiltonian, ${\cal H}_1$. We provide explicit recursion relations for the 
components of the eigenvectors and 
for the eigenvalues. 
  We also show how it is related to the $S-$gap Lam\'e  
equation, using the coherent states of $SU(2)$.

In section~\ref{numerics} we provide numerical evidence that the spectrum
of the AH Hamiltonian  may indeed be meaningfully approximated by that of 
${\cal H}_1$. 

We end with our conclusions and a discussion of  directions of further 
inquiry. 

\section{The Quantum Group $sl_q(2)$ and the AH Hamiltonian}\label{notation}
The Azbel--Hofstadter (AH) Hamiltonian is a tight-binding model for a
 single Bloch electron on a two dimensional lattice and in the 
presence of an external, homogeneous and transverse magnetic 
field~\cite{harper,azbel,zak,chambers,hofst,wannier,wz}. 

The AH Hamiltonian is 
\begin{equation}
\label{awhham}
{\cal H}_{\mathrm AH}=\sum_{<m,n>}t_{nm}e^{{\mathrm i}A_{nm}}c_n^{\dagger}c_m
\end{equation}
where $t_{nm}$ are the hopping amplitudes, $c_n^{\dagger}$, $c_n$ creation and
annihilation operators for the electron at site $n=(n_x,n_y)\in {\bf Z}\times {\bf Z}$ and  $A_{nm}$ the line element, {\em viz.} 
\begin{equation}
\label{linelem}
\frac{e}{c}\int_{n}^{m}{\bf A\cdot}d{\bf x}=A_{nm}
\end{equation}
such that, through  each elementary plaquette, we have a flux $\Phi$,
\begin{equation}
\label{flux}
\prod_{\mathrm plaquette}e^{{\mathrm i}A_{nm}}=e^{{\mathrm i}\Phi/\Phi_0}
\end{equation}
where $\phi\equiv\Phi/\Phi_0=2\pi M/N$\, with $M$ and $N$ coprime integers and $\Phi_0$ the flux quantum. 

For such values of the magnetic flux the Hilbert space becomes an 
infinite number of identical $N-$dimensional copies,
 due to the existence of an infinite number of magnetic translations,
 that commute with Hamiltonian ${\cal H}_{\mathrm AH}$. 
In the case of a square lattice the dynamics is restricted to a 
$N\times N$ square, magnetic, lattice. In the Landau gauge,
\begin{equation}
\label{landau}
\begin{array}{lr}
A_x\equiv A_{n_x,n_x+1}=0 & A_y\equiv A_{n_y,n_y+1}=\Phi\cdot n_x\\
\end{array}
\end{equation}
the Bloch wavefunction $\psi(n_x,n_y)=\exp({\mathrm i}(k_xn_x+k_yn_y))\psi_{k}(n_x)$, where $(k_x,k_y)\in [0,2\pi)\times[0,2\pi)$ and 
\begin{equation}
\label{blochwf}  
\psi_k(n_x+N)=\psi_k(n_x)
\end{equation}
satisfies Harper's equation 
\begin{equation}
\label{harper}
t_x(e^{{\mathrm i}k_x}\psi_{n+1}+e^{-{\mathrm i}k_x}\psi_{n-1})+
t_y(e^{{\mathrm i}(k_y+n\phi)}+e^{-{\mathrm i}(k_y+n\phi)})\psi_n=E\psi_n
\end{equation}
where $t_x,t_y$ are the hopping parameters and $\psi_n\equiv\psi_k(n)$. 

We shall restrict our discussion to the case of an isotropic square lattice, 
$t_x=t_y=1$. 

Harper's equation can be written in matrix form
\begin{equation}
\label{matharper}
H\Psi=E\Psi
\end{equation}
where $\Psi=(\psi_1,\ldots,\psi_N)^T$ and 
\begin{equation}
\label{matHam}
H=e^{{\mathrm i}k_x}P+e^{-{\mathrm i}k_x}P^{-1}+e^{{\mathrm i}k_y}Q+
                                                e^{-{\mathrm i}k_y}Q^{-1}
\end{equation}
The matrices $Q$ and $P$ are
\begin{equation}
\label{QandP}
\begin{array}{c}
Q_{k,l}=\omega^{(k-1)}\delta_{k,l}\\
\\
P_{k,l}=\delta_{k-1,l},\,\,\,k,l=1,\ldots,N
\end{array}
\end{equation}
with $\omega=\exp(2\pi{\mathrm i}M/N)$ and all operations are performed mod $N$. 

The matrices $P$ and $Q$ generate the magnetic translations and the finite Heisenberg group through the Weyl commutation relation
\begin{equation}
\label{weylCR}
QP=\omega PQ
\end{equation}
The Heisenberg group elements 
\begin{equation}
\label{heisenelem}
{\cal J}_{r,s}=\omega^{r\cdot s/2}P^rQ^s
\end{equation}
provide a projective representation of the two dimensional translation group
 mod $N$~\cite{weyl,zachfair}
\begin{equation}
\label{projtrans}
{\cal J}_{r,s}{\cal J}_{r',s'}=\omega^{(r' \cdot s-r\cdot s')/2}{\cal J}_{r+r',s+s'}
\end{equation}
The factor $1/2$ in the exponents of the previous relations is defined as $s+1$ for $N=2s+1$ and $s$ an integer, while, for $N$ even, it is defined as $0.5$ . 
The matrices ${\cal J}_{r,s}$ are traceless, unitary and they have period $N$, {\em viz.}
\begin{equation}
\label{Jmat}
\begin{array}{c}
{\cal J}_{r,s}^{\dagger}={\cal J}_{-r,-s}\\
\\
{\cal J}_{r,s}^{N}={\bf 1}_{N}\\
\end{array}
\end{equation}
The matrices $Q$ and $P$ are related through the finite Fourier transform matrix $F$,
\begin{equation}
\label{fourier}
F_{k,l}=\frac{1}{\sqrt{N}}\omega^{(k-1)(l-1)},\,\,\,k,l=1,\ldots,N
\end{equation}
by
\begin{equation}
\label{QPrel}
F\cdot P=Q\cdot F
\end{equation}
Because of the symmtery 
\begin{equation}
\label{symmetry}
\begin{array}{c}
{\cal J}_{r,s}^{\dagger}\cdot P\cdot{\cal J}_{r,s}=\omega^{-s}P\\
\\
{\cal J}_{r,s}^{\dagger}\cdot Q\cdot{\cal J}_{r,s}=\omega^{r}Q\\  
\end{array}
\end{equation}
the Brillouin torus $[0,2\pi)\times[0,2\pi)$ is reduced to a smaller one
$[0,2\pi l/N)\times[0,2\pi l/N)$,where $l=M$ mod $N$ (recall that the flux is 
$2\pi M/N$). 

In ref.~\cite{wz} Wiegmann and Zabrodin made the important observation that, for the midpoint of the Brillouin zone,
\begin{equation}
\label{midpoint}
\cos N k_x +\cos N k_y=0
\end{equation}
the spectrum of ${\cal H}_{\mathrm AH}$ is determined by the roots $z_m$ of polynomials $P(z)$ of degree $N-1$, that interpolate the wavefunction
\begin{equation}
\label{polwave}
\begin{array}{c}
\psi_n=P(q_0^n),\,\,\,n=0,\ldots,N-1 \\
\\
q_0=e^{{\mathrm i}\pi M/N}\\
\end{array}
\end{equation}
The energy spectrum is given by 
\begin{equation}
\label{energyspectrum}
E={\mathrm i}q_0^n(q_0-q_0^{-1})\sum_{m=1}^{N-1}z_m
\end{equation}
and the roots $z_m$ satisfy the algebraic Bethe {\em Ansatz} equations
\begin{equation}
\label{betheansatz}
\frac{q_0-z_k^2}{q_0z_k^2-1}=\prod_{m\neq k}\frac{q_0z_k-z_m}{z_k-q_0z_m}
\end{equation}
where $k=1,\ldots,N-1$. 

Behind these findings is the quantum group $sl_{q_0}(2)$, which is a deformation of the Lie algebra $sl(2)$~\cite{jimbo}
\begin{equation}
\label{slq2}
\begin{array}{c}
q_0^{J_0}J_{\pm}q_0^{-J_0}=q_0^{\pm 1}J_{\pm }\\
\\
\left[ J_{+},J_{-}\right]=\displaystyle\frac{q_0^{2J_0}-q_0^{-2J_0}}{q_0-q_0^{-1}}\\
\end{array}
\end{equation}
where $J_{\pm},\,J_0$ are expressed in terms of the magnetic translations ${\cal J}_{r,s}$ and the Hamiltonian, in a specific, ``quasi-Landau'' gauge, is 
expressed, in terms of the $J_{\pm}$ as
\begin{equation}
\label{slq2ham}
{\cal H}_{\mathrm AH}={\mathrm i}(q_0-q_0^{-1})(J_{-}\pm J_{+})
\end{equation}
at the midpoint(s) of the Brillouin zone. 

In ref.~\cite{fadkas} Faddeev and Kashaev 
observed that the AH Hamiltonian is equivalent to the three site chiral 
Potts model~\cite{potts}, whose quantum group symmetry is known to 
lead to the algebraic Bethe {\em Ansatz} equations over 
specific Riemann surfaces. 
Thus they were able to generalize the result of Wiegmann and Zabrodin to 
arbitrary points of the Brillouin zone, as 
well as to anisotropic square and triangular lattices.

\section{The Cartesian deformation of $sl(2)$ and the AH Hamiltonian}\label{slq2C}

Some years ago a new deformation of the $sl(2)$ algebra, in the Cartesian basis,
 was proposed, which has a cyclic symmetry for the generators~\cite{wfzhed}
\begin{equation}
\label{cartesian}
\begin{array}{lcr}
qJ_1J_2-q^{-1}J_2J_1&=&(q^2-q^{-2})J_3\\
qJ_2J_3-q^{-1}J_3J_2&=&(q^2-q^{-2})J_1\\
qJ_3J_1-q^{-1}J_1J_3&=&(q^2-q^{-2})J_2\\
\end{array}
\end{equation}
The representation theory for real values of the deformation $q$ was studied
by Fairlie~\cite{wfzhed} and, in more detail, by  Zhedanov {\em et al.}, 
who pointed out that this algebra generates the properties of a class of 
Askey-Wilson polynomials. Recently in ref.~\cite{spiridonov} 
the representations for $q$ a primitive root of unity of order $N$ were 
constructed and classified. It is known that, in this case, the 
standard (``Jimbo'') deformation of $sl(2)$ has irreducible representations of
all dimensions smaller or equal to the order of the root. These representations
depend on three complex parameters and are of two types: type A irreps have 
a classical $sl(2)$ analog, while type B are cyclic (i.e. not ladder) of order 
$N$~\cite{roche-arnaudon}.
  
The Cartesian algebra, $sl_q^{C}(2)$ has one Casimir element 
\begin{equation}
\label{Casimir}
C_2=(q+q^{-1})(J_1^2+J_2^2)-\{J_3,\tilde{J}_3\}
\end{equation}
where $(q^2-q^{-2})\tilde{J}_3=q^{-1}J_1J_2-qJ_2J_1$. In ref.~\cite{bi,af,afn}, 
using results from Finite Quantum Mechanics, it was realized that the
 AH Hamiltonian,at any point of the Brillouin zone, can be written as the anticommutator of the two operators $J_1$ and $J_2$,
\begin{equation}
\label{hamafawh}
{\cal H}_{\mathrm AH}=\frac{1}{q+q^{-1}}\{J_1,J_2\}
\end{equation}
where $J_1$,$J_2$ and $J_3$ realize an $N-$dimensional representation of
 $sl_{q}^{C}(2)$ and may be expressed in terms of the generators of the 
Heisenberg group, $J_{r,s}$ through
\begin{equation}
\label{cartheis}
\begin{array}{lcr}
J_1&=&e^{{\mathrm i}\sigma}J_{m,-m}+e^{-{\mathrm i}\sigma}J_{-m,m}\\
\\
J_2&=&e^{{\mathrm i}\rho}J_{m,m}+e^{-{\mathrm i}\rho}J_{-m,-m}\\
\\
J_3&=&e^{{\mathrm i}k_y}Q+e^{-{\mathrm i}k_y}Q^{-1}\\
\end{array}
\end{equation}
where $q=\omega^{m^2}$, $\omega=e^{2\pi{\mathrm i}/N}$,$m\equiv (1/2)$mod $N=(N+1)/2$ and $\sigma=(k_x-k_y)/2$, $\rho=(k_x+k_y)/2$. 

The unitary representation of $sl_q^C(2)$ constructed above is 
irreducible with Casimir element 
\begin{equation}
\label{casmirslq2c}
C_2=8\cos\frac{2\pi M}{N}m
\end{equation}
and 
\begin{equation}
\label{tildeJ3slq2c}
\tilde{J}_3=e^{{\mathrm i}k_x}P+e^{-{\mathrm i}k_x}P^{-1}
\end{equation}
The generators $J_1$, $J_2$, $J_3$ can be cyclically permuted at the 
Brillouin point $(k_x,k_y)=(0,0)$ by three $N\times N$ unitary matrices, $U_1$, $U_2$, $U_3$ 
\begin{equation}
\label{U1U2U3}
\begin{array}{c}
U_1^{-1}J_2U_1=J_3\\
\\
U_2^{-1}J_3U_2=J_1\\
\\
U_3^{-1}J_1U_3=J_2\\
\end{array}
\end{equation}
Theses matrices leave the corresponding $J_i$'s invariant. This situation 
brings to mind the rotation group in three dimensions, where $\pi/2-$rotations 
around the coordinate axes  cyclically permute the generators of the group. 
In the appendix~\ref{Us} we construct explicitly the group generated by 
$U_1$, $U_2$, $U_3$. 

Before concluding this section, we establish the relation of the 
$N-$dimensional unitary matrices $J_1$, $J_2$, $J_3$ with matrices 
$J_{\pm}$ and $J_0$, which satisfy the (Jimbo) deformation of $sl(2)$, i.e.
 $sl_q(2)$. 

It is straightforward to see that, defining $J_{\pm}$ and $J_0$ by 
\begin{equation}
\label{reljimheis}
\begin{array}{c}
J_{\pm}=\displaystyle\pm\frac{1}{q-q^{-1}}\left({\cal J}_{\pm m,\pm m}+{\cal J}_{\pm m,\mp m}\right)\\
\\
Q=\omega^{mJ_0}\\
\end{array}
\end{equation}
the Jimbo deformed algebra is satisfied. Then, introducing matrices
\begin{equation}
{\cal T}_{\pm}=P^{-1}J_{\pm}P
\end{equation}
we establish the relation between $J_1$, $J_2$, $J_3$ and $J_{\pm}$ and $J_0$ as
\begin{equation}
\begin{array}{ccc}
J_1&=&\displaystyle\frac{1}{q+q^{-1}}\left(q^2J_{+}+q^{-2}J_{-}-
{\cal T}_{+}-{\cal T}_{-}\right)\\
\\
 J_{2}&=&\displaystyle -\frac{1}{q+q^{-1}}\left(q^{-2}J_{+}+q^2J_{-}-
{\cal T}_{+}-{\cal T}_{-}\right)\\
\\
 J_3&=&\displaystyle q^{2J_0}+q^{-2J_0}\\
\end{array}
\end{equation}
where $q=\omega^{m^2}$. 

Finally, we note that the constructed representation for $J_{\pm}$ and $J_0$ is
 of the cyclic type~\cite{roche-arnaudon} since we can check that eq.~(\ref{reljimheis}) implies 
\begin{equation}
J_{+}^N=2\times {\bf 1}_N
\end{equation}

\section{The eigenvalue problem for the AH Hamiltonian}\label{su2qsym}
Although in the literature the eigenvalue problem has been discussed in 
various contexts, we present here a compact method for the determination of 
the eigenvalues and eigenvectors, using the tridiagonal form of the AH
Hamiltonian. This method is especially suited for fast numerical calculations 
for large $N$. 

The eigenvalue problem, in components, is the following
\begin{equation}
\label{evprob}
e^{{\mathrm i}k_y}\psi_{k-1}+2\cos\left((k-1)\frac{2\pi}{N}+k_x\right)\psi_k+
e^{-{\mathrm i}k_y}\psi_{k+1} = E\psi_k
\end{equation}
where $k=1,\ldots,N$ and we use periodic boundary conditions
\begin{equation}
\psi_{N+l} = \psi_{l}
\end{equation}
for all $l$. We define homogeneous variables $z_k$ 
\begin{equation}
z_k = \frac{\psi_{k+1}}{\psi_k}
\end{equation}
and we use the ``M\"obius'' notation
\begin{equation}
\left(\begin{array}{cc} a & b\\ c & d\\ \end{array}\right)\cdot z\equiv
\frac{az+b}{cz+d}
\end{equation}
to rewrite the equations~(\ref{evprob}) in the form 
\begin{equation}
\label{moebevp}
z_k=
\left(\begin{array}{cc} E-2\cos\left((k-1)\frac{2\pi}{N}+k_x\right) & 
-e^{{\mathrm i}k_y}\\ e^{-{\mathrm i}k_y} & 0\\ \end{array}\right)\cdot 
z_{k-1}
\end{equation}
for $k=1,\ldots,N$. The last equation, $k=N$, gives, by iteration, the relation
\begin{equation}
\label{lasteq}
z_N = \left\{\prod_{k=1}^{N-1}
\left(\begin{array}{cc} E-2\cos\left((k-1)\frac{2\pi}{N}+k_x\right) & 
-e^{{\mathrm i}k_y}\\ e^{-{\mathrm i}k_y} & 0\\ \end{array}\right)\right\}\cdot
z_0
\end{equation}
Since $z_N=z_0$, we deduce the characteristic equation, {\em viz.}
\begin{equation}
\label{chareq}
{\mathrm det}\left[
\prod_{k=1}^{N-1}
\left(\begin{array}{cc} E-2\cos\left((k-1)\frac{2\pi}{N}+k_x\right) & 
-e^{{\mathrm i}k_y}\\ e^{-{\mathrm i}k_y} & 0\\ \end{array}\right) -{\bf 1}_{2\times 2}\right]=0
\end{equation}
and we may then compute the components of the corresponding eigenvector, by 
recursion, starting from $z_N$, which is a fixed point of the M\"obius 
transformation, eq.~(\ref{moebevp}). 

It is known~\cite{chambers} that 
the characteristic polynomial takes the form
\begin{equation}
\label{chambers}
P_N(E)+(-)^N\times  4\left(\cos Nk_x +\cos Nk_y\right)
\end{equation}
where $P_N(E)$ is a polynomial of degree $N$ in $E$ and coefficients that
 do not depend on $(k_x,k_y)$. The study of the structure of the gaps between 
the eigenvalues is a very interesting problem in functional analysis and 
noncommutative geometry (cf. ref~\cite{bellissard-crete} for a recent review) 

It should be noted that the existence of a zeromode for eq.~(\ref{chareq})
  depends on $(k_x,k_y)$. 
We observe that, for $N$ even and $r=N/2$, we obtain 
\begin{equation}
\label{even}
{\cal J}_{r,r}^{\dagger}{\cal H}_{\mathrm AH}{\cal J}_{r,r}+
{\cal H}_{\mathrm AH}=0
\end{equation}
for any point of the Brillouin zone.
This implies the existence of a reflection symmetry for the spectrum, i.e. for any eigenvalue $E$ there exists an eigenvalue $-E$. This does not necessarily imply the existence of the eigenvalue $E=0$; if it exists, however, it must necessarily be doubly degenerate.

For $N$ odd the reflection symmetry, $E\leftrightarrow -E$ is approximate. 

On the other hand, for $k_x+k_y=\pi$, the Hamiltonian anticommutes with the 
discrete Fourier transform, which implies that, for any $N$, the reflection symmetry is realized and, for $N$ odd, these points belong to the midband, eq.~(\ref{midpoint}) and the zeromode always exists. 

In the context of the quantum group $U_q(sl(2))$, the eigenvector corresponding 
to $E=0$ is a $q-$Askey--Wilson polynomial~\cite{wz} and the structure of 
its roots as a function of the magnetic flux has been studied by Kohmoto {\em 
et al.}~\cite{kohmoto}.

\section{The symmetries of an Euler top}\label{symmetries}
We introduce now the classical analog  Hamiltonian
\begin{equation}
\label{su2}
{\cal H}_1 = \frac{1}{2\mathrm i}\left\{{\cal S}_1,{\cal S}_2\right\} = 
\frac{1}{4\mathrm i}({\cal S}_{+}^2-{\cal S}_{-}^2)
\end{equation}
in the spin-$S$ representation, with $Q=2S+1$, where ${\cal S}_i$ are the 
standard $SU(2)$ generators.    

We propose to study the eigenvalue problem of ${\cal H}_1$ with the hope to 
gain some intuition, which may be useful for the real problem (i.e. that of the 
AH Hamiltonian). 
   
The matrix elements of ${\cal H}_1$ are 
\begin{equation}
\label{ham}
({\cal H}_1)_{k,k^{'}}= \\ 
{\displaystyle \frac{1}{4\mathrm i}}
\left( {\sf a}_{k}\delta_{k^{'},k+2} - {\sf a}_{k^{'}}
\delta_{k^{'},k-2}\right)
\end{equation}
where
\begin{equation}
\label{eleme}
{\sf a}_{k}\equiv a_{k-S-1}a_{k-S}
\end{equation}
and
\begin{equation}
a_m=\sqrt{S(S+1)-m(m+1)}
\end{equation} 
with $k,\,k^{'}=1,\ldots,2S+1$.  
The structure of the matrix shows immediately that the even-numbered components
decouple from the odd-numbered ones for any value of $S$. 

The three operators
 ${\cal R}_{1,2}\equiv {\mathrm e}^{-{\mathrm i}\pi{\cal S}_{1,2}}$ and
${\cal R}_3\equiv {\mathrm e}^{{\mathrm i}\frac{\pi}{2}{\cal S}_3}$.  
will prove useful for the decomposition of the 
$2S+1$--dimensional eigenspace in convenient subspaces. In components they read as follows
\begin{eqnarray}
\label{3R}
({\cal R}_{1})_{k,k^{'}} &=& {\mathrm e}^{{\mathrm i}\pi S}\delta_{k+k^{'},2S+2}\nonumber\\
({\cal R}_{2})_{k,k^{'}} &=& {\mathrm e}^{2{\mathrm i}\pi S}(-1)^{k-1}
\delta_{k+k^{'},2S+2}\nonumber\\
({\cal R}_{3})_{k,k^{'}} &=& {\mathrm e}^{{\mathrm i}\frac{\pi}{2} S}
({-\mathrm i})^{k-1}\delta_{k,k^{'}}
\end{eqnarray} 
They satisfy the 
following commutation relations
\begin{eqnarray}
\label{commuteR}
{\cal R}_1{\cal R}_2 &=& (-1)^{2S}{\cal R}_2{\cal R}_1=({\cal R}_3)^2\nonumber\\
{\cal R}_3{\cal R}_1 &=& {\cal R}_2{\cal R}_3\nonumber\\
({\cal R}_{1,2})^2 &=& ({\cal R}_3)^4 = (-1)^{2S}
\end{eqnarray}
From the above it follows that ${\cal R}_{1,2,3}$ anticommute with the Hamiltonian and 
we can construct two projectors, that {\em commute} with 
the Hamiltonian,
\begin{equation}
\label{projectp}
{\cal P}_{\pm}=\frac{1}{2}\left(1\pm {\mathrm e}^{-{\mathrm i}\pi S}
({\cal R}_3)^2\right)  
\end{equation}
and
\begin{equation}
\label{projectq}
{\cal Q}_{\pm}=\frac{1}{2}\left(1\pm {\mathrm e}^{-{\mathrm i}\pi S}
{\cal R}_{3}{\cal R}_{2}\right)
\end{equation}
${\cal P}_{\pm}$ project on the subspaces of the 
odd-- and even--indexed components of the eigenvectors, 
while ${\cal Q}_{\pm}$ on the positive and negative energy eigenspaces. 
The Hamiltonian  ${\cal H}_1$ 
thus may be written as a direct sum of Hamiltonians ${\cal H}_{\pm}$
\begin{equation}
{\cal H}={\cal H}_{+}\oplus{\cal H}_{-}
\end{equation}
where 
\begin{equation}
{\cal H}_{\pm}={\cal H}{\cal P}_{\pm}
\end{equation}
and the dimensions of the corresponding Hilbert spaces are, for integer $S$,  
$S+1$ and $S$ respectively while, for half-integer $S$,  they both have dimension $S+
{\displaystyle\frac{1}{2}}$. 
In components
\begin{eqnarray}
({\cal H}_{+})_{k,k^{'}} &=& {\sf a}_{2k+1}\delta_{k^{'},k+1}
                                          -{\sf a}_{2k-1}\delta_{k^{'},k-1}
   \nonumber\\
({\cal H}_{-})_{k,k^{'}}&=& {\sf a}_{2k+2}\delta_{k^{'},k+1}
                                          -{\sf a}_{2k}\delta_{k^{'},k-1}
\end{eqnarray}
It is straightforward to numerically diagonalize the Hamiltonians ${\cal H}_{\pm}$ and 
compute the eigenvalues and  eigenfunctions. 
The anticommutation relations~(\ref{commuteR}) imply 
a spectrum antisymmetric about $E=0$ for any value of $S$ and for 
both subspaces. The general structure of the eigenvectors is qualitatively 
similar to that of the one-dimensional harmonic oscillator and their 
phase structure (real/imaginary components appear symmetrically or 
antisymmetrically up to factors of $\pm 1$ or $\pm {\mathrm i}$) may 
be deduced from the antidiagonal structure of 
the projectors ${\cal Q}_{\pm}$.   
Analytically, it is 
possible to completely describe the $E=0$ case and provide recursion relations
for the $E\neq 0$ cases. 

We introduce the standard notation for the M\"obius transformation 
\begin{equation}
\label{moebius}
\left(\begin{array}{cc}
                 a & b \\
                 c & d \\
        \end{array}
          \right)\cdot w \equiv \frac{aw+b}{cw+d}
\end{equation}
It is easy to show that the eigenvalue problem
${\cal H}|\Psi\rangle=E|\Psi\rangle$ takes the following form in component 
notation (where ${\cal E}\equiv 2{\mathrm i}E$)
\begin{eqnarray}
\label{compon}
-{\sf a}_{2k-1}\psi_{2k-1}+{\sf a}_{2k+1}\psi_{2k+3} &=& {\cal E}\psi_{2k+1}
\nonumber\\
-{\sf a}_{2k}\psi_{2k}+{\sf a}_{2k+2}\psi_{2k+4} &=& {\cal E}\psi_{2k+2}
\end{eqnarray}
corresponding to the odd and even subspaces. Defining 
\begin{eqnarray}
z_k\equiv\frac{\psi_{2k+1}}{\psi_{2k-1}}\nonumber\\
w_k\equiv\frac{\psi_{2k+2}}{\psi_{2k}}
\end{eqnarray}
we solve  eqs.~(\ref{compon}) by the M\"obius transformation
\begin{equation}
\label{componodd}
z_{k+1}=\left\{\prod_{m=1}^{k}\left(
                          \begin{array}{cc}
				{\cal E} & {\sf a}_{2m-1}\\
                              {\sf a}_{2m+1} & 0\\
                           \end{array}
                               \right)\right\}\cdot z_1
\end{equation}
where $k=1,\ldots,S-1$ and $z_1={\cal E}/{\sf a}_1$. The last equation in this 
subspace takes the form
\begin{equation}
\label{chodd}
0 = \left(\begin{array}{cc}
	{\cal E}_{-} & {\sf a}_{2S-1}\\
        1 & 0\\
          \end{array}\right)\cdot z_{S}
\end{equation}
which is the characteristic equation for $E$ in the odd-numbered sector. 
Similarly, for the even components one has ($w_1={\cal E}/{\sf a}_2$)
\begin{equation}
\label{componev}
w_{k+1}=\left\{\prod_{m=1}^{k}\left(
			\begin{array}{cc}
				{\cal E} & {\sf a}_{2m}\\
				{\sf a}_{2m+2} & 0\\
			\end{array}\right)\right\}\cdot w_1
\end{equation}
and the corresponding characteristic equation
\begin{equation}
\label{chev}
0=\left(\begin{array}{cc}
	{\cal E}_{+} & {\sf a}_{2S-2}\\
	1 & 0\\
	\end{array}\right)\cdot w_{S-1}
\end{equation}
The above analysis becomes explicit for $E=0$. There are several possibilities,
depending on the values of $S$. 
\begin{itemize}
\item integer {\em S}: \\
$E=0$ is an eigenvalue only for $S=4k+1,4k+3$ (non-degenerate) and $S=4k+2$ (doubly 
degenerate). 

The explicit form of the corrresponding eigenvector  
$|\Psi_{E=0}^{b}\rangle\equiv (\psi_1,\ldots,\psi_{2S+1})$ is
\begin{equation}
\label{bosonic}
\psi_{4n+2}=\frac{\prod_{m=0}^{n-1}{\sf a}_{4m+2}}{\prod_{m=1}^{n}{\sf a}_{4m}}\psi_2
\end{equation}
for the non-degenerate case and identical to that of the half-integer $S$ 
(see below) for the doubly degenerate case; all other components are zero.

\item half-integer {\em S}: \\
There are two possible cases to consider: $S=(4k+1)/2$ and $S=(4k+3)/2$. 
In the first, $E=0$ is an eigenvalue, that is doubly degenerate, one
 belonging to the $(+)$, the other to the $(-)$ subspace. 

The two eigenvectors, corresponding to $E=0$ are $|\Psi^{(1)}\rangle\equiv
(\psi_1^{(1)},\ldots,\psi_{2S+1}^{(1)})$ and 
$|\Psi^{(2)}\rangle\equiv (\psi_1^{(2)},\ldots,\psi_{2S+1}^{(2)})$ 
\begin{eqnarray}
\label{fermionic}
\psi_{4n+1}^{(1)}=\frac{\prod_{m=0}^{n-1}{\sf a}_{4m+1}}
{\prod_{m=1}^{n}{\sf a}_{4m-1}}\psi_1^{(1)}
\nonumber\\ 
\psi_{4n+2}^{(2)}=\frac{\prod_{m=0}^{n-1}{\sf a}_{4m+2}}
{\prod_{m=1}^{n}{\sf a}_{4m}}\psi_2^{(2)} 
\end{eqnarray}
For $S=(4k+3)/2\,\,\,\,E=0$ doesn't belong to the spectrum, as may be proved by
explicit calculation. 
\end{itemize}
In the above relations the $\psi_1,\psi_2$ are normalization constants. 
To determine the components of the eigenvectors ($z_k,w_k$) one has simply 
to evaluate the two-dimensional matrix products in 
eqs.~(\ref{componodd},\ref{componev}) for each root ${\cal E}$ of the characteristic
equations~(\ref{chodd},\ref{chev}).

Another approach to the problem is provided by the coherent state 
representation of $SU(2)$ where we write the generators as differential 
operators and the algebraic eigenvalue problem becomes a Schr\"odinger 
problem in a class of potentials. This connection has been used in the 
inverse way for classifying potentials that have a ``quasi-integrable''
spectrum, i.e. a finite number of eigenstates decouples from the rest and may 
be determined by {\em finite } matrix methods~\cite{zasul,turbi,turbi1}.  
Even more recently, these same equations have appeared in the study of 
the correlation functions of Wess-Zumino-Witten models 
on the torus~\cite{etingof}. 
 
In fact ${\cal H}_1$ belongs to a family of Hamiltonians with the same 
spectrum
\begin{eqnarray}
\label{gen}
{\cal H}={\mathrm e}^{{\mathrm i}\alpha {\cal S}_3}{\cal H}_1{\mathrm e}^
{-{\mathrm i}\alpha {\cal S}_3}=\nonumber\\
\cos 2\alpha\cdot{\cal H}_1+\sin 2\alpha\cdot{\cal H}_2
\end{eqnarray}
where ${\cal H}_1=\{{\cal S}_1,{\cal S}_2\}$ and ${\cal H}_2=
{\cal S}_1^2-{\cal S}_2^2$. ${\cal H}_2$ is known to describe 
isotropic paramagnets in two dimensions and a similar Hamiltonian has been 
studied in ref.~\cite{zasul}. Indeed, by a ${\cal S}_2$ rotation of angle 
$\pi/2$, one obtains the Zaslavskii--Ul'yanov Hamiltonian ${\cal H}_3$~\cite{zasul} in zero external magnetic field     
\begin{equation}
\label{diffham}
{\cal H}_3={\mathrm e}^{-{\mathrm i}\frac{\pi}{2}{\cal S}_2}
           {\mathrm e}^{-{\mathrm i}\frac{\pi}{2}{\cal S}_3}
		{\cal H}_1
	   {\mathrm e}^{{\mathrm i}\frac{\pi}{2}{\cal S}_3}
	   {\mathrm e}^{{\mathrm i}\frac{\pi}{2}{\cal S}_2} = {\cal S}_3^2-{\cal S}_2^2
\end{equation}

In the coherent state basis, the $SU(2)$ generators have the following form  
\begin{eqnarray}
\label{diffop}
{\cal S}_1 & = & S\cos\phi-\sin\phi{\displaystyle\frac{d}{d\phi}}\nonumber\\
{\cal S}_2 & = & S\sin\phi+\cos\phi{\displaystyle\frac{d}{d\phi}}\\
{\cal S}_3 & = & -{\mathrm i}{\displaystyle\frac{d}{d\phi}}\nonumber
\end{eqnarray}
and the eigenvalue problem for ${\cal H}_3$ is
\begin{equation}
\label{coherent}
\left[(1+\cos^2\phi)\frac{d^2}{d\phi^2}+(S-\frac{1}{2})\sin 2\phi\frac{d}{d\phi}+(E+S^2\sin^2\phi+S\cos^2\phi)\right]\Phi(\phi) = 0
\end{equation}
The components $\psi_{m} \,m=-S,\ldots,S$ of  eigenvectors of 
${\cal H}_3$, are 
related to the function $\Phi(\phi)$ by 
\begin{equation}
\label{mateq}
\Phi(\phi)=\sum_{m=-S}^{S}\frac{\psi_m}{\sqrt{(S-m)!(S+m)!}}
{\mathrm e}^{{\mathrm i}m\phi}
\end{equation}
If we change variables, following ref.~\cite{zasul},  from 
$(\phi,\Phi(\phi))$ to $(x,\Psi(x))$, defined by
\begin{eqnarray}
\label{lamevar}
\Psi(x)&=&\Phi(\phi(x))(1+\cos^2\phi(x))^{-S/2}\nonumber\\
\frac{d}{dx} &=& \frac{1}{\sqrt{2}}(1+\cos^2\phi(x))^{1/2}\frac{d}{d\phi}
\end{eqnarray}
eq.~(\ref{coherent}), after a redefinition
\begin{equation}
\label{redef}
x=u-{\bf K}(1/2)
\end{equation}
where ${\bf K}(1/2)$ is the complete elliptic integral of the first 
kind~\cite{erdelyi},
 becomes the  $S-$gap Lam\'e equation~\cite{erdelyi}
\begin{equation}
\label{lame}
\left[\frac{d^2}{du^2}+\frac{E+S(S+1)}{2}-S(S+1)\frac{{\mathrm s\mathrm n}^2 u}{2}\right]
\Psi(u)=0
\end{equation}
where ${\mathrm s\mathrm n} u$ is the elliptic sine of modulus $1/\sqrt{2}$~
\cite{erdelyi}.
This equation has polynomial solutions in terms of 
elliptic sines and cosines and appears in the inverse scattering method 
for the KdV equation and the potential corresponds to the soliton 
solutions of the KdV.

\section{Numerical Results}\label{numerics}
In this section we compute numerically the eigenvalues and eigenvectors 
of the AH Hamiltonian and the model Hamiltonian, ${\cal H}_1$, for various 
dimensions. The size of the matrix representation of the AH Hamiltonian depends 
solely on the denominator of the magnetic flux, $\phi=2\pi M/N$, where $M$ and
$N$ are relatively prime integers. In the case of ${\cal H}_1$, the size is 
$N=2S+1$ and, as noted previously, there are two decoupled,  invariant subspaces, of 
dimension $S$ and $S+1$ respectively.  For simplicity, we shall compare the
 spectra, for $N$ odd,  
of ${\cal H}_{\mathrm AH}$ and 
${\cal H}_1$. 

Furthermore, our toy Hamiltonian does not depend on the Bloch momenta 
$(k_x,k_y)$ but, since the Brillouin zone has size equal to the flux, our 
approximation should be better, for flux $2\pi/N$, at the point
 $(k_x,k_y)=(0,0)$, where the AH Hamiltonian has maximal symmetry (commutes with
the finite Fourier transform matrix) and has both odd and even eigenfunctions, 
as is the case for ${\cal H}_1$. 

\begin{figure}[thp]
\epsffile{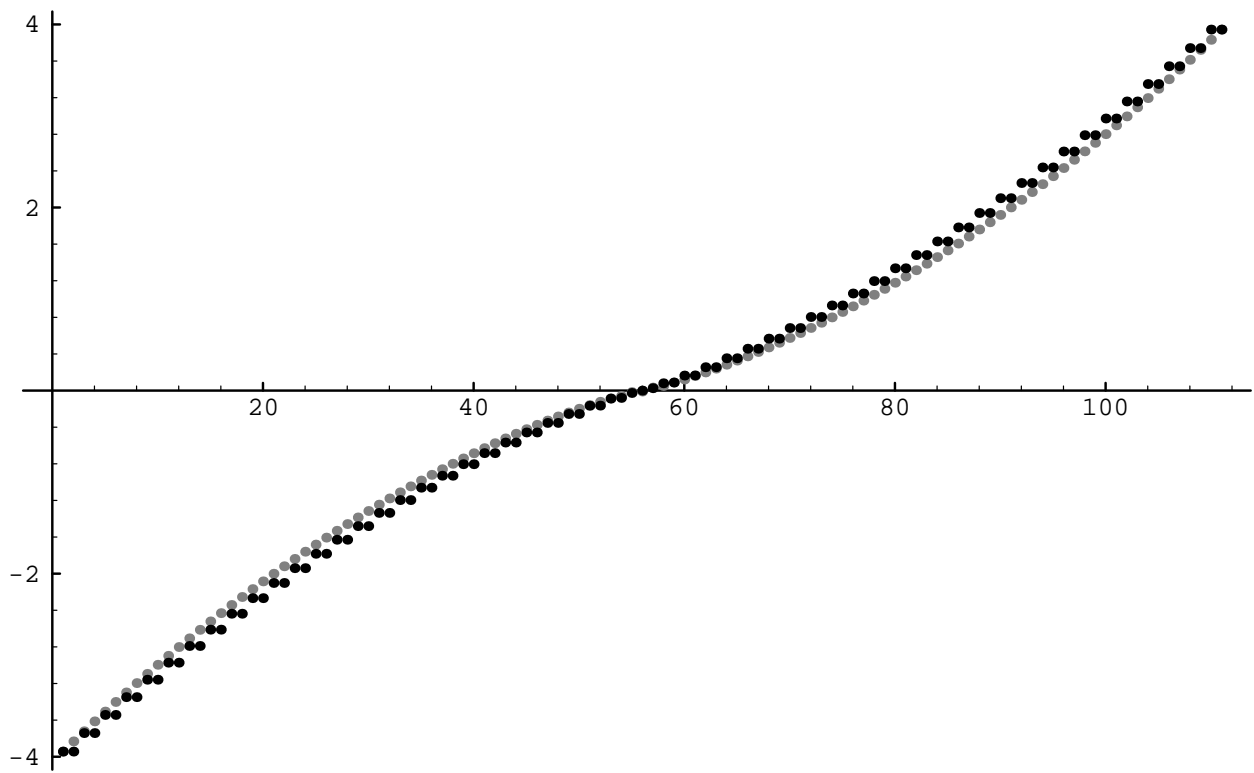}
\caption[]{Comparison of the spectra of ${\cal H}_{\mathrm AH}$ (grey points)
and ${\cal H}_1$ (black points), $N=111$.}
\label{one} 
\end{figure}

\begin{figure}[thp]
\epsffile{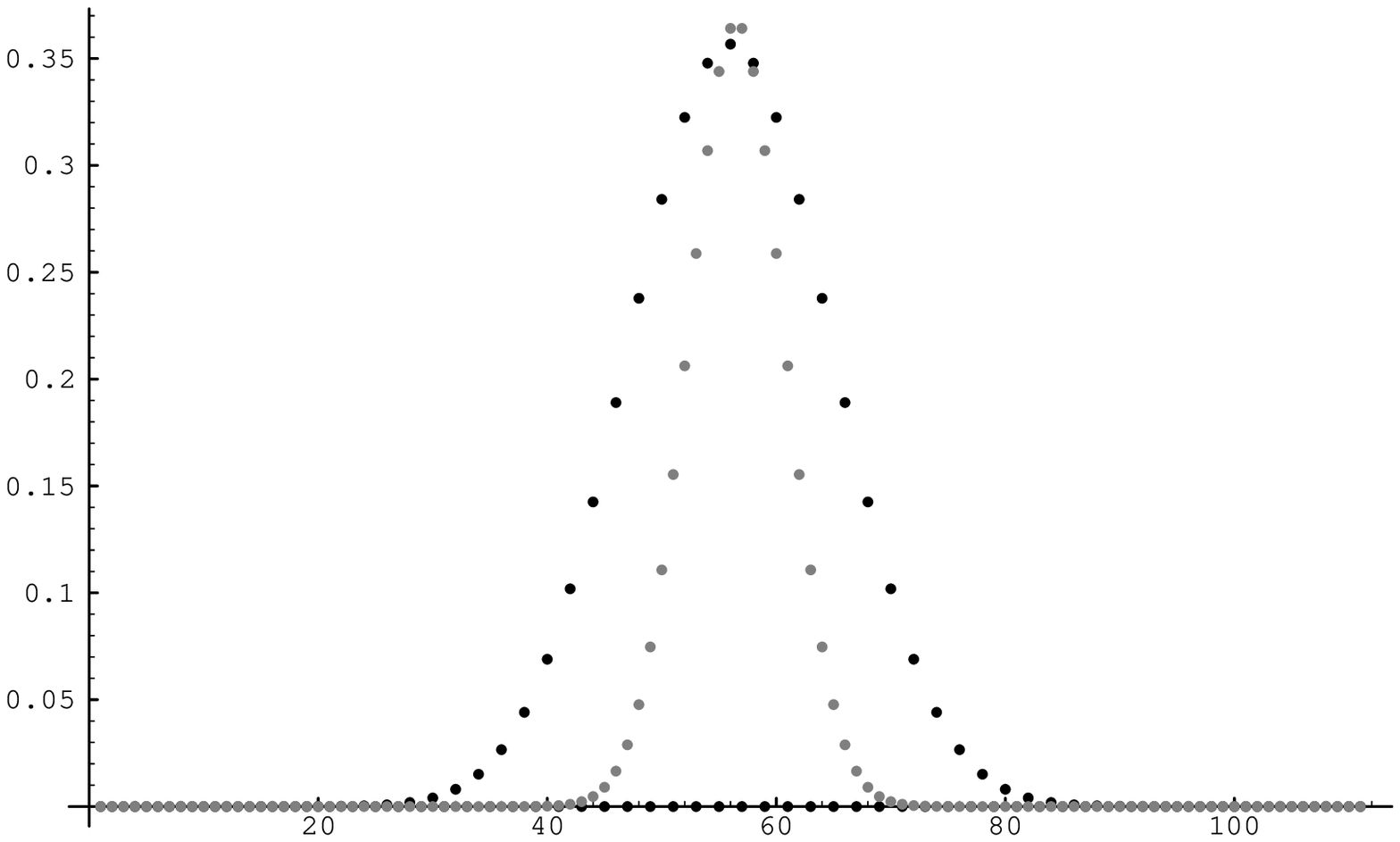}
\caption[]{Comparison of the ground state eigenvectors of ${\cal H}_{\mathrm AH}$ (grey points)
 and ${\cal H}_1$ (black points), $N=111$.}
\label{two}
\end{figure}

\begin{figure}[thp]
\epsffile{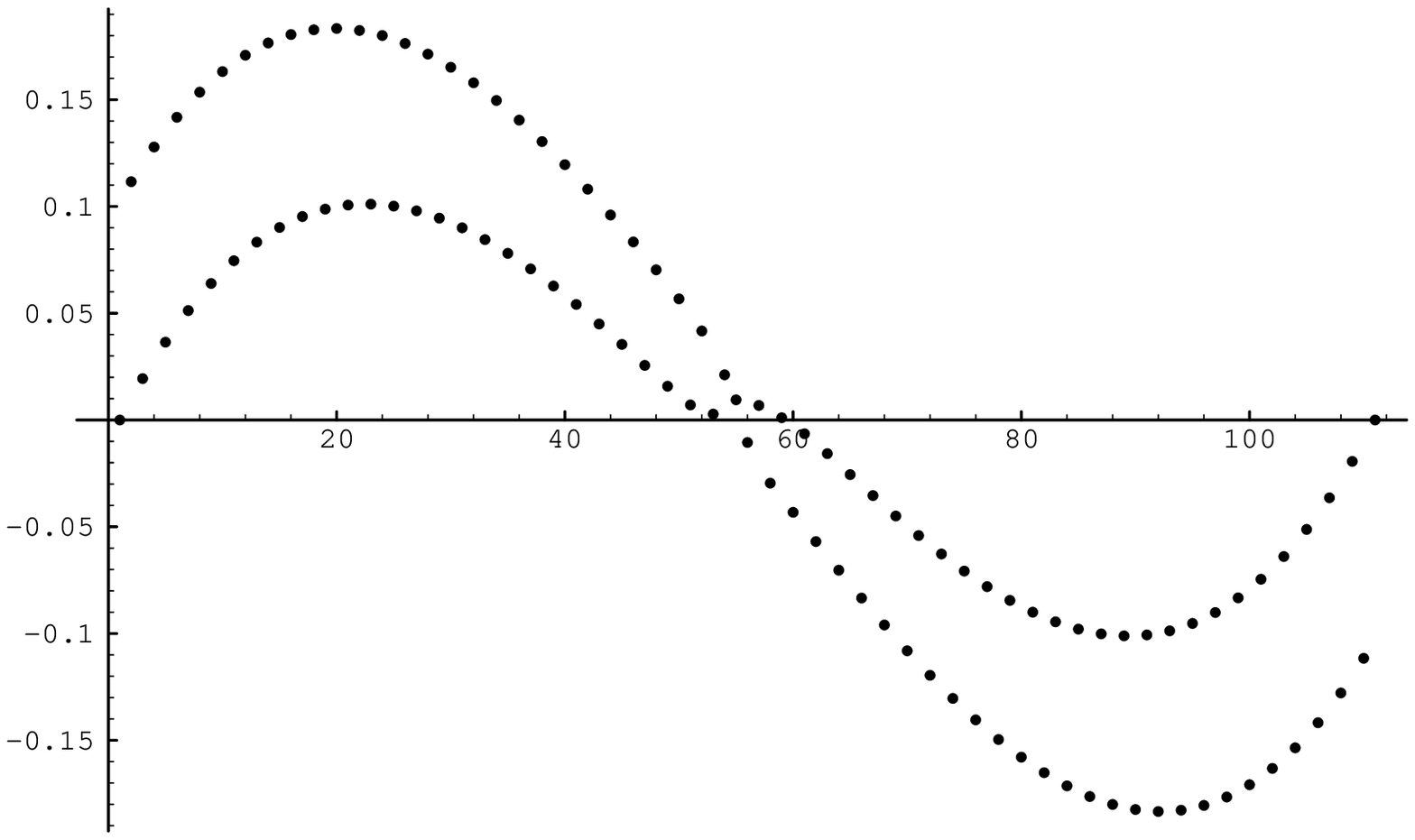}
\caption[]{Plot of $E_{{\mathrm AH}}(n)-E_{{\cal H}1}(n)\times 
(E_{{\mathrm AH}}(1)/E_{{\cal H}1}(1))$. Note the striking difference between 
the values of the even and the odd subspaces.}
\label{three}
\end{figure}

\begin{figure}[thp]
\epsffile{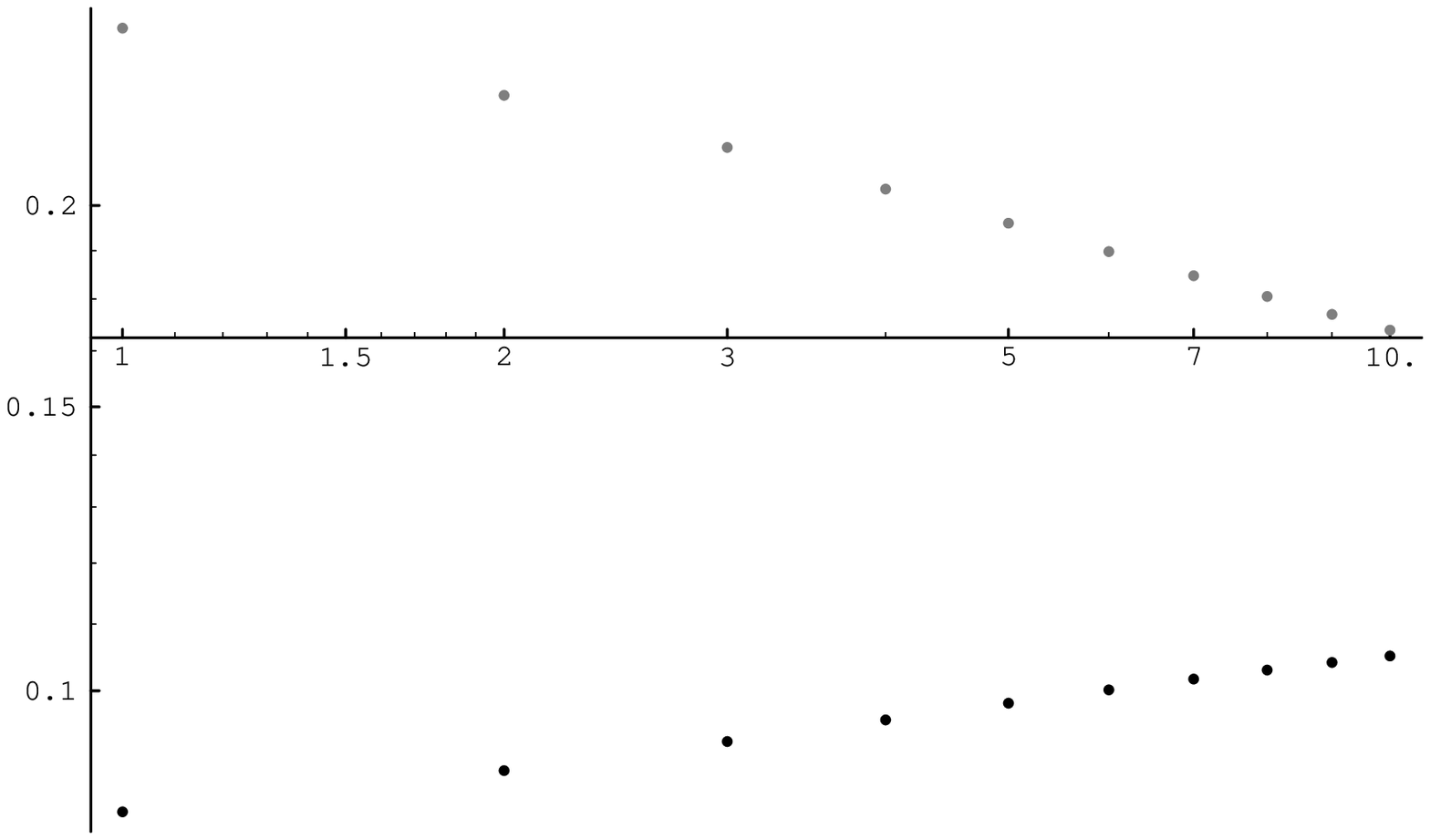}
\caption[]{Log-Log plot for the maximum difference for $N=55$ to $N=145$ in steps of 10 for the 
spectra $E_{{\mathrm AH}}$ and $E_{{\cal H}1}$. 
 Black dots are for the even subspaces and gray dots for the odd 
subspaces of ${\cal H}_1$.}
\label{four}
\end{figure}

In fig.~\ref{one} we compare the spectra for the case $N=111$ and in 
figure~\ref{two} we present the ground state eigenvectors for the two Hamiltonians. 
In comparing the spectra we have normalized them both to have the same numerical 
value for the ground state energy. 
In fig~\ref{three} we give the normalized difference of the two spectra for 
$N=111$. In fig.~\ref{four} we display the maximum error 
for several values of $N$. 
\section{Conclusions and Perspectives}\label{perspectives}
The main point of this paper is the striking similarity between the spectrum
of the AH Hamiltonian and that of our Euler top model. For the moment 
this is an empirical observation, for which we still lack a physical 
appreciation from first principles. 

From the quantum group symmetry point of view the cyclic character of the 
$U_q(sl(2))$ representations, which appear in this problem would forbid 
a direct classical large $N$ limit. However, our numerical results indicate 
that our approximation, through the Euler top, gets better with increasing $N$. 
This probably suggests some kind of ``analytic continuation'' between the 
the Cartesian deformation of $SU(2)$ and the ``classical'' $SU(2)$ for 
the root of unity case. This is still an open problem, for which our results 
seem to indicate a promising direction of attack.  

Another problem is the dependence on the Brillouin zone parameters $(k_x,k_y)$ 
as well as on the flux $\phi=2\pi M/N$; indeed we recall that our results 
hold only for $M=1$ and it is an interesting question, for instance, to 
calculate the ``Hofstadter butterfly'' for the Euler top. 

We hope to return to these questions in future work. 

{\sl Acknowledgements}: This research was partially supported by the EU grant
SCI--0430--C. 
We would also like to thank the staff of the Laboratoire 
de Physique Th\'eorique at the Ecole Normale Sup\'erieure for their warm 
hospitality.

\appendix
\section{Quantum Rotations}\label{Us}
In order to construct the three matrices $U_1$, $U_2$ and $U_3$, mentioned in section~\ref{slq2C} we have to determine the inner automorphism group of the discrete Heisenberg group matrices ${\cal J}_{r,s}$, i.e. matrices $U(A)$, such that 
\begin{equation}
U^{\dagger}(A){\cal J}_{r,s}U(A)=J_{(r,s)A}
\end{equation}
where $r,\,s=0,\ldots,N-1$, for every matrix $A\in SL(2,{\bf Z}_N)$~\cite{bi, af}.
 It is obvious, from the definition of the generators $J_1$, $J_2$ and $J_3$,
 that we can construct matrices $U_1$, $U_2$ and $U_3$, if we find three $2\times 2$ matrices $A_1$, $A_2$ and $A_3$, which leave invariant the set of indices$(m,-m)$, $(m,m)$, $(0,1)$ respectively. We check immediately that there exist three abelian subgroups of $SL(2,{\bf Z}_N)$, which do the job, generated by
\begin{equation}
\begin{array}{c}
A_1=\left(\begin{array}{cc} m & m \\ -m & 3m\\ \end{array}\right)\\
A_2=\left(\begin{array}{cc} 3m & m \\ -m & m\\ \end{array}\right)\\
A_3=\left(\begin{array}{cc} 1 & 2 \\ 0 & 1 \\ \end{array}\right)\\
\end{array}
\end{equation}
We set $U_i=U(A_i)$, with $i=1,2,3$. Using the explicit forms for $U(A),\,A\in SL(2,{\bf Z}_N)$ from ref.~\cite{af}, we find 
\begin{equation}
\label{Umat}
\begin{array}{ccc}
(U_1)_{k,l}&=&\frac{1}{\sqrt{N}}\omega^{m(l-k)(3l-k-2)}\times
\left\{\begin{array}{c}1\\-{\mathrm i}\\ \end{array}\right\} \\
(U_2)_{k,l}&=&\frac{1}{\sqrt{N}}\omega^{m(l-k)(l-3k+2)}\times
\left\{\begin{array}{c}1\\-{\mathrm i}\\ \end{array}\right\} \\
(U_3)_{k,l}&=&\delta_{k,l}\omega^{m(k-1)^2}\\
\end{array}
\end{equation}
where $k,l=1,\ldots,N$ and the symbol 
\begin{equation}
\left\{\begin{array}{c}1\\-{\mathrm i}\\ \end{array}\right\}\equiv\left\{
\begin{array}{l}
1\,\,N\equiv\,1\,{\mathrm mod}\,4\\
-{\mathrm i},\,\,N\equiv\,-1\,{\mathrm mod}\,4\\
\end{array}
\right.
\end{equation}
We note finally that the cyclic groups generated by $U_1$,$U_2$ and $U_3$ are of order $N$. 
The matrices $U_1$, $U_2$ and $U_3$ can be used to define specific discrete Askey-Wilson polynomials as the columns of the matrix
\begin{equation}
{\cal P}_{\mathrm AW}=U_1^{\dagger}U_2
\end{equation}
cf. the papers of Zhedanov and collaborators in ref.~\cite{wfzhed}. 

\end{document}